\documentstyle[amssymb,thmsa,12pt,sw20lart]{article}
\input{tcilatex}
\begin{document}

\author{O. Bohigas$^{1}$ and M. P. Pato$^{2}$ \\
%EndAName
$^{1}$Laboratoire de Physique Th\'{e}orique et Mod\`{e}les Statistique, \\
B\^{a}timent 100, Universit\'{e} Paris-Sud, \\
91405 Orsay Cedex, France\\
$^{2}$Instituto de Fisica, Universidade de Sao Paulo\\
C.P.20516,01498 Sao Paulo, S.P., Brazil}
\title{Missing levels in correlated spectra}
\maketitle
\date{}

\begin{abstract}
Complete spectroscopy (measurements of a complete sequence of consecutive
levels) is often considered as a prerequisite to extract fluctuation
properties of spectra. It is shown how this goal can be achieved even if
only a fraction of levels are observed. The case of levels behaving as
eigenvalues of random matrices, of current interest in nuclear physics, is
worked out in detail.
\end{abstract}

It is by now well established that, at least with present statistical
significance limited by the relatively scarce amount of high quality data
available, the statistical properties of compound nucleus resonances with
fixed quantum numbers are consistent with the predictions of the
Wigner-Dyson random matrix model (Gaussian Orthogonal Ensemble, GOE)\cite
{Haq}. It is presently understood, not in the form of mathematical theorems
but rather from numerical experiments and from theoretical insight based on
semiclassical approaches, that the origin of this (universal) behavior has
to be found in the chaotic character of the underlying classical motion\cite
{Giannoni}. Analyses have been mainly performed for two degrees of freedom
systems (one particle in a two-dimensional box), which can be also
experimentally exemplified in studying resonances of microwave cavities with
adequate boundaries or of quartz or metallic blocks (elastodynamics) (see,
for instance, contributions by A. Richter, H. J. St\"{o}ckmann and C.
Ellegaard\cite{Aberg}).

Efforts have also been devoted to find out from which excitation energy on,
starting from the ground state (where, by analyzing the systematics of
ground state energies, evidence for the coexistence between regular and
chaotic motion has been given\cite{Leboeuf}) the `random matrix theory'
behavior holds\cite{Egidy}. This is a difficult task because, below neutron
threshold, one can hardly be sure that all levels are detected. Work is also
currently performed to investigate the statistical properties of the fine
structure of analog or other giant resonant states, at excitation energies
which may be below or far above the neutron threshold\cite{Richter}.

At different degrees, most of these studies must face the difficulty of
establishing the extent to which the sequences of energies analyzed are
complete ( no missing levels) as well as how the missing levels may distort
the statistical properties of the otherwise ideal (complete) sequence. It is
the purpose of this letter to present some results which may be useful in
this context.

The most ambitious goal may be stated as the one of detecting the location
of, say, one missing level on an otherwise complete sequence. Dyson, in a
recent review\cite{Dyson}, uses information theory concepts and argues that
correlations in a sequence may provide the necessary redundancy from which
error correcting codes can be constructed. At one extreme where no
correlations and therefore no redundancy are present (Poissonian sequence),
there is no possibility of detecting one missing level. At the other
extreme, a sequence of equally spaced levels (picket fence ), there is a
maximum redundancy and a missed level can be obviously detected as a hole in
the spectrum. Eigenvalues of random matrices, which exhibit characteristic
correlations, correspond to an intermediate situation between these two
extremes. The attempts to locate in the last case a single missed level have
remained unsuccessful so far. However it should be mentioned that for
two-dimensional chaotic systems where, besides correlations of the order of
one mean spacing as described by random matrices, the presence and the role
of long range correlations governed by the shortest periodic orbits and
reflected in Weyl's law describing the average spectral density, is well
understood. It is then possible to approximately locate, from the study of
spectral fluctuations, a single missed level\cite{Schmit}.

Here we address a less ambitious question, namely to study how the spectral
fluctuations of an ideally complete sequence are affected when a fraction $f$
$(0<f<1)$ of levels are detected. It will be assumed that the sequence is
infinite and stationary and that the levels are dropped at random from the
complete sequence (random sampling). The last assumption is presumably
justified when observing compound nucleus resonances with limited detection
sensitivity. Indeed, in random matrix theory level positions and widths are
uncorrelated, a feature which is confirmed by the analysis of experimental
data.

We start considering the $n$-point correlation functions $R_{n}\left(
E_{1},E_{2},...,E_{n}\right) $ that give the joint probability density of
finding levels located around each of the values $E_{1},E_{2},...,E_{n}$.
For the incomplete sequence these functions keep their form and they are
only reduced by a factor $f^{n}$ since $f$ is the probability of observing
one given level. We can therefore write the relation

\begin{equation}
r_{n}\left( E_{1},E_{2},...,E_{n}\right) =f^{n}R_{n}\left(
E_{1},E_{2},...,E_{n}\right) ,  \label{40}
\end{equation}
where small cases denote the actual observed quantities and capital letters
those of the complete spectrum. We shall keep this notation in what follows,
namely use the same letter, capital or small case, to refer to corresponding
quantities for complete or incomplete sequences, respectively. For the
spectral density (\ref{40}) gives

\begin{equation}
r_{1}\left( E\right) =fR_{1}\left( E\right) ,  \label{41}
\end{equation}
and, from (\ref{41}), the unfolding procedure that maps the levels onto a
new spectrum whose density is equal to one yields the relation

\begin{equation}
x_{k}=\int_{-\infty }^{E_{k}}r_{1}\left( E\right) dE=fX_{k},  \label{42}
\end{equation}
where $k=1,2,...$ labels the levels which are observed. The normalization
will always be kept the same (the mean spacing $<s>=1$ for the complete as
for incomplete sequences). This is in contrast to some of the expressions
given in \cite{Mitchell}.

One is usually interested in the $n$-point cluster functions $Y_{n}\left(
X_{1},X_{2},...,X_{n}\right) $ from which lower order correlations are
subtracted. From their definition\cite{Mehta} and (\ref{40}) we deduce

\begin{equation}
y_{n}\left( x_{1},x_{2},...,x_{n}\right) =Y_{n}\left( \frac{x_{1}}{f},\frac{%
x_{2}}{f},...,\frac{x_{n}}{f}\right) .  \label{46}
\end{equation}
These important equations show that the correlation functions keep their
form, the only modification is a rescaling of variables. Thus, if the
fraction $f$ is known, the correlation functions can be reconstructed from
the partially observed spectrum.

It is well known that the number variance $\Sigma ^{2}\left( L\right) $
(variance of the number of levels contained in an interval of length $L$)
and the Dyson-Mehta $\Delta $-statistics (least square deviation, in an
interval of length $L$, of the actual staircase counting function from its
linear approximation) are related to the two-point cluster function through%
\cite{Mehta}

\begin{equation}
\Sigma ^{2}\left( L\right) =L-2\int_{0}^{L}\left( L-x\right) Y_{2}\left(
x\right) dx,  \label{60}
\end{equation}

\[
\Delta \left( L\right) =\frac{L}{15}-\frac{1}{15L^{4}}\int_{0}^{L}\left(
L-x\right) ^{3}\left( 2L^{2}-9xL-3x^{2}\right) Y_{2}\left( x\right) dx. 
\]
From (\ref{46}) it follows immediately that

\begin{equation}
\sigma ^{2}\left( L\right) =\left( 1-f\right) L+f^{2}\Sigma ^{2}\left( \frac{
L}{f}\right)  \label{64}
\end{equation}
and

\begin{equation}
\delta \left( L\right) =\left( 1-f\right) \frac{L}{15}+f^{2}\Delta \left( 
\frac{L}{f}\right) .  \label{64a}
\end{equation}
Notice the appearance of a linear term in both equations, if $f<1.$

Another set of statistical measures are the $E(n,s)$ functions that give the
probabilities of finding $n$ levels, with $n=0,1,2...,$ inside an interval
of length $s.$ If for the complete spectrum they are given by $E(n,s),$ then
their expressions if only a fraction $f$ of levels is observed is

\begin{equation}
e(n,s)=\sum_{k=n}^{\infty }\frac{k!}{n!\left( k-n\right) !}f^{n}\left(
1-f\right) ^{k-n}E(k,\frac{s}{f}).  \label{4}
\end{equation}
This follows from the fact that $1-f$ is the probability that one level is
missed, $f$ that it is not and $\frac{k!}{n!\left( k-n\right) !}$ counts the
number of ways $k-n$ points can be removed from $k$ points. From $e(n,s),$
the spacing distributions $p(n,s),$ the probability density of finding $n$
levels ($n=0,1,2...)$ between two given levels separated by a distance $s,$
can be derived using the relation

\begin{equation}
P(n,s)=\frac{d^{2}}{ds^{2}}\sum_{k=0}^{n}\left( n-k+1\right) E(k,s);
\label{16}
\end{equation}
one obtains

\begin{equation}
p(n,s)=\sum_{k=n}^{\infty }\frac{k!}{n!\left( k-n\right) !}f^{n}\left(
1-f\right) ^{k-n}P(k,\frac{s}{f}).  \label{20}
\end{equation}
For the nearest neighbor distribution (NND), i.e. $p(s)=p(0,s)$ (\ref{20})
becomes

\begin{equation}
p(s)=\sum_{k=0}^{\infty }\left( 1-f\right) ^{k}P(k,\frac{s}{f}).  \label{28}
\end{equation}
The equations derived above imply that the general relation

\[
y_{2}\left( x\right) =1-\sum_{n=0}^{\infty }p\left( n,x\right) 
\]
is fulfilled, as it should.

Expression (\ref{28}) was first proposed as an ${\it {anzatz}}$ in ref. \cite
{Bilpuch}, and in ref. \cite{Mitchell} it has been shown that the
coefficients $f\left( 1-f\right) ^{k}$ maximize the Shannon entropy with
constraints appropriately defined. Here it is obtained in a direct way.

We first apply, as a check, the previous equations to a Poissonian spectrum
of uncorrelated levels in which case they must remain invariant (they should
not depend on $f$). If, for example, we substitute in Eq. (\ref{4}) the
expression

\begin{equation}
E(n,s)=\frac{s^{n}}{n!}\exp (-s)  \label{76}
\end{equation}
for the $E$-functions of a Poisson sequence, one obtains

\begin{equation}
e(n,s)=\sum_{k=n}^{\infty }\frac{k!}{n!\left( k-n\right) !}f^{n}\left(
1-f\right) ^{k-n}\frac{1}{n!}\left( \frac{s}{f}\right) ^{n}\exp (-\frac{s}{f}%
)=E(n,s).  \label{77}
\end{equation}
It follows that any other measure derived from the $E(n,s)$ functions will
also remain invariant. Since $Y_{2}\left( X_{1},X_{2},...,X_{n}\right) =0,$
the n-point cluster functions of the incomplete sequence also vanish and all
other measures derived from them also remain invariant.

Before considering GOE spectra with a fraction of observed levels, let us
first recall some of the GOE expressions ($f=1$) to be used\cite{Mehta}. The
two-point cluster function reads

\begin{equation}
Y_{2}\left( x\right) =\left[ \frac{\sin (\pi x)}{\pi x}\right] ^{2}-\left[
Si\left( \pi x\right) -\pi \epsilon \left( x\right) \right] \left[ \frac{%
\cos (\pi x)}{\pi x}-\frac{\sin (\pi x)}{\left( \pi x\right) ^{2}}\right] ,
\label{77a}
\end{equation}
where $Si\left( \pi x\right) $ is the sine-integral and $\epsilon \left(
x\right) =\pm \frac{1}{2}$ if $x>0$ or $x<0$ and $\epsilon \left( 0\right) =0
$ if $x=0.$ The number variance $\Sigma ^{2}(L)$ is given by

\begin{equation}
\Sigma ^{2}(L)=\frac{2}{\pi ^{2}}\left[ \ln (2\pi L)+\gamma +1\right] -\frac{%
1}{4}+O\left( L^{-1}\right) ,  \label{77b}
\end{equation}
where $\gamma =0.577...$ is Euler's constant and the $\Delta $-statistics is
given by

\begin{equation}
\Delta (L)=\frac{1}{\pi ^{2}}\left[ \ln (2\pi L)+\gamma +\frac{5}{4}\right] -%
\frac{1}{8}+O\left( L^{-1}\right) .  \label{77c}
\end{equation}
The NND distribution is well approximated by the Wigner surmise ($2$x$2$
matrices)

\begin{equation}
P(s)=P(0,s)=\frac{\pi }{2}s\exp \left( -\frac{\pi }{4}s^{2}\right) .
\label{78}
\end{equation}
The next to nearest neighbor distribution $P(1,s)$ is given by the NND of
the symplectic ensemble which again is well approximated by the $2$x$2$
matrix result

\begin{equation}
P(1,s)=\frac{8}{3\pi ^{3}}(\frac{4}{3})^{5}s^{4}\exp \left( -\frac{16s^{2}}{%
9\pi }\right)  \label{78a}
\end{equation}
(care has been take to insure $\left\langle s\right\rangle =2$). The higher (%
$k=2,3,...$) spacing distributions $P(k,s)$ are well approximated by their
(Gaussian) asymptotic form, centered at $k+1$ and variances $V\left(
k\right) $ given by\cite{Fre78}

\begin{equation}
V^{2}\left( k\right) \simeq \Sigma ^{2}(L=k)-\frac{1}{6}.  \label{78b}
\end{equation}

We apply now the above expressions to GOE spectra when only a fraction $f$
of levels is observed. On Fig 1 is displayed for different values of $f$ the
function $Y_{2}$, illustrating the scaling behavior Eq. (\ref{46}) using (%
\ref{77a}).

By comparing Eq. (\ref{64}) with (\ref{77b})and (\ref{64a}) with (\ref{77c})
one immediately sees one major effect of the incompleteness of a sequence.
Instead of a logarithmic increase of the number variance or of the
Dyson-Mehta statistic ($f=1$), one has a linear increase for $f<1$ of slope $%
\left( 1-f\right) $ or $\left( 1-f\right) /15$ respectively.

Consider finally $p\left( s\right) $, the NND very often discussed in the
literature. It is given by (\ref{28}), (\ref{78}), (\ref{78a}), (\ref{78b}).
The range of the argument determines the number of terms to be included in (%
\ref{28}). The slope at the origin given by the first term in (\ref{28}) is
increased by a factor $1/f$ with respect to the GOE value. Let us also
mention that, in fact, Eq. (\ref{28}) was already used to analyze high
quality data of $^{238}$U neutron resonances\cite{Liou} and of $^{48}$Ti
proton resonances\cite{Mitchell}. It was concluded that a fraction of about
10\% of the resonances is missing in both cases and that the correction is
important in investigating the parity dependence of nuclear level densities%
\cite{Undraa}.

It is convenient, in particular to exhibit the asymptotic behavior of $%
p\left( s\right) $, to rewrite Eq. (\ref{28}) in a different form. Let us
first separate the first $K$ terms in the sum and approximate the rest (the
infinite sum from $k=K$ to $k=\infty $) by an integral in which $k$ is
treated as a continuous variable

\begin{equation}
p\left( s\right) =\sum_{k=0}^{k=K-1}\left( 1-f\right) ^{k}P(k,\frac{s}{f}%
)+\int_{K}^{\infty }\frac{dk}{\sqrt{2\pi V^{2}(k)}}\exp \left[ k\ln (1-f)-%
\frac{1}{2V^{2}(k)}(\frac{s}{f}-k-1)^{2}\right] .  \label{79a}
\end{equation}
A steepest descent approximation to the integral can be worked out by
mapping the exponent in the integrand onto a parabola as

\begin{equation}
F(k)=-k\ln (1-f)+\frac{1}{2V^{2}(k)}(\frac{s}{f}-k-1)^{2}=F(k_{s})+t^{2},
\label{79b}
\end{equation}
where $k_{s}$, the stationary point, is the root of the equation $F^{\prime
}(k)=0$. This transcendent equation can be solved approximately if one
considers the variance $V^{2}(k)$ as a slowly varying function of its
argument, in which case one can write

\begin{equation}
k_{s}=\frac{s}{f}-1+V^{2}(\frac{s}{f}-1)\ln (1-f).  \label{79e}
\end{equation}
Replacing then the coefficient $\frac{1}{\sqrt{2\pi V^{2}(k)}}\frac{dk}{dt}$
by its value at the stationary point, we obtain the closed expression

\begin{equation}
p(s)=\sum_{k=0}^{k=K-1}\left( 1-f\right) ^{k}P(k,\frac{s}{f})+\frac{1}{2}%
\exp \left[ -F(k_{s})\right] \left[ 1-\text{erf}(t_{s})\right] ,  \label{79c}
\end{equation}
where

\begin{equation}
t_{s}=\left\{ 
\begin{array}{c}
\sqrt{F(K)-F(k_{s})}\text{ for }k_{s}<K \\ 
-\sqrt{F(K)-F(k_{s})}\text{ for }k_{s}>K
\end{array}
\right.  \label{79d}
\end{equation}
In the asymptotic region, $\frac{s}{f}$ large, (\ref{79c}) takes the simple
expression

\begin{equation}
p(s)\sim \frac{1}{1-f}\exp \left[ \frac{s}{f}\ln (1-f)\right]  \label{79m}
\end{equation}
illustrating that the spacing distribution approaches a Poisson distribution
as $f$ tends to zero. We have therefore a family of NND distributions,
parametrized by the fraction $f$ of the observed levels, that interpolates
between the GOE and the Poisson statistics.

In Fig 2 comparison is made with numerical simulations. It can be seen that
for $f=0.90$ the asymptotic regime Eq. (\ref{79m}) is not yet reached in the
range considered whereas for $f=0.50$ it is already reached for $s\succsim 2$%
.

In conclusion, we have shown how the problem of missing, at random, a
fraction of levels of a correlated spectrum can be solved. From the
observation of an incomplete spectrum the results presented here have a
twofold application: i) if the missing fraction is known, one can recover
the statistical properties of the complete spectrum, ii) if the statistical
nature of the complete spectrum is known, an estimate of the fraction of
missed levels can be obtained.

Fruitful discussions with A. Heine, G. E. Mitchell and A. Richter are
acknowledged. This work is supported by a project CAPES-COFECUB. M.P.P. is
partially supported by the Brazilian agency Conselho Nacional de Pesquisas
(CNPq).

{\bf Figure Captions}

Fig. 1 The two-level cluster function $Y_{2\text{ }}.$ Full line: Eqs. (\ref
{46}) and (\ref{77a}). Points: obtained with numerical simulation for
different values of the fraction $f$ of levels observed.

Fig. 2 Nearest neighbor spacing distribution $p\left( s\right) $ for two
values of $f$ . Full line: theory Eq. (\ref{28}) or (\ref{79c}). Dashed
line: Wigner surmise Eq. (\ref{78}). Dotted line: asymptotic behavior as
given by Eq. (\ref{79m}). Histograms: numerical simulation.

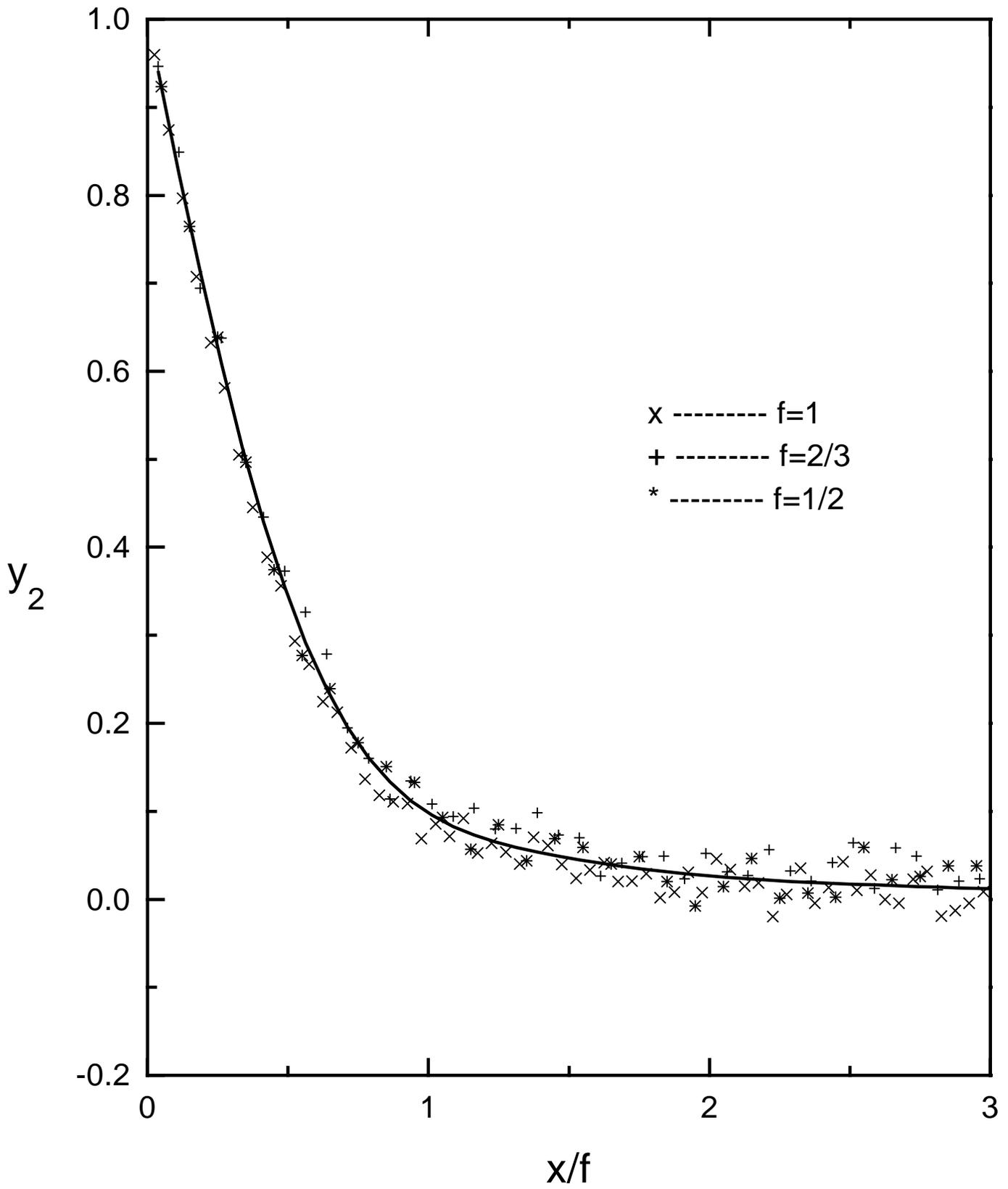
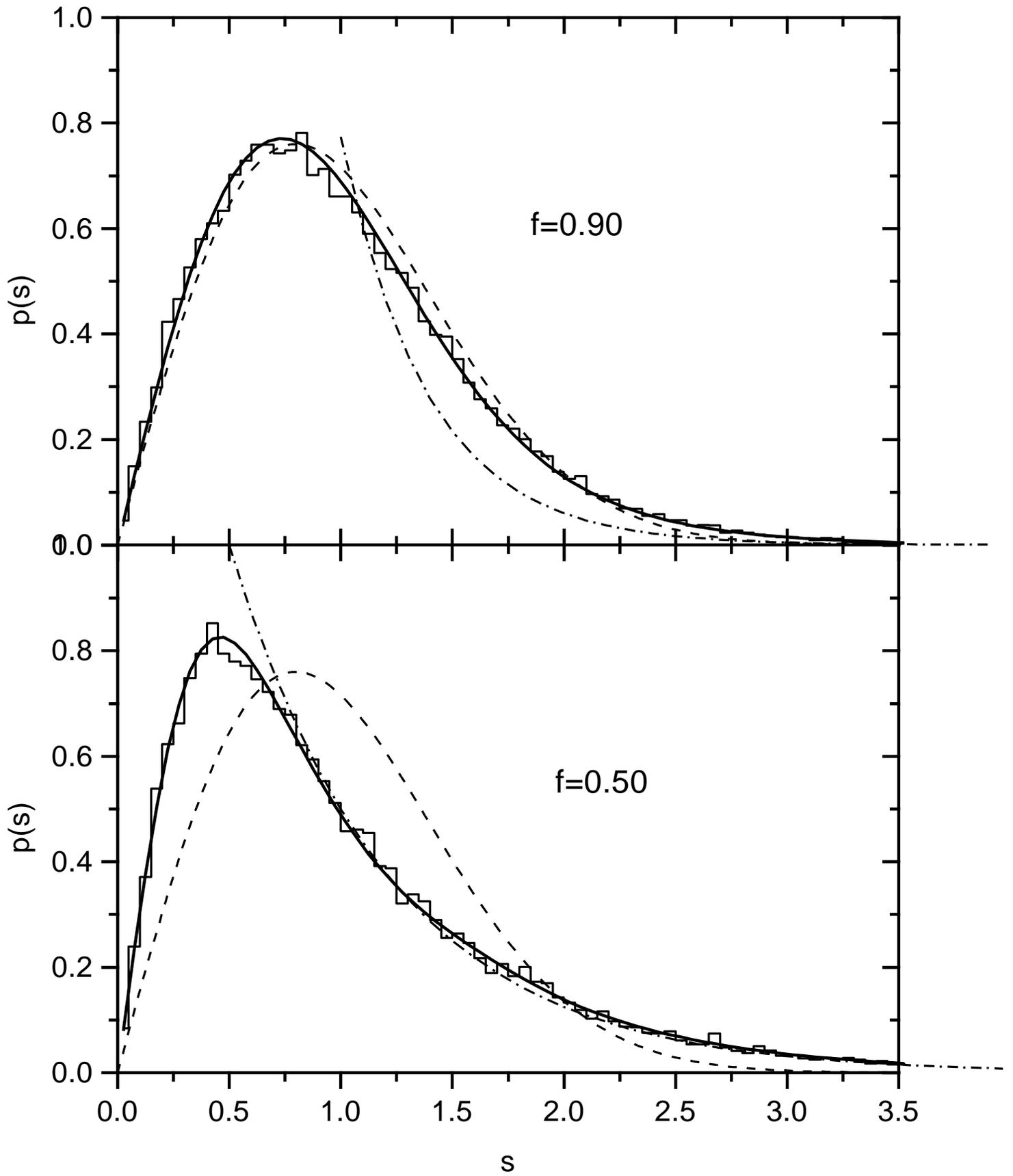
\end{document}